\documentclass[%
reprint,
amsmath,amssymb,
aps,
]{revtex4-1}

\usepackage{color}
\usepackage{graphicx}
\usepackage{dcolumn}
\usepackage{bm}
\usepackage[pdftex,plainpages=false,colorlinks=true,linkcolor=blue, citecolor=blue, urlcolor=blue]{hyperref}

\begin{document}
	
	\preprint{APS/123-QED}
	
	\title{Spin-Wave Excitations in $\mathrm{La_3Ni_2O_7}$: A Dynamical Mean-Field Theory \\ and Bethe-Salpeter Equation Study
	}

	\author{Zhihui Luo}
	\affiliation{Guangdong Provincial Key Laboratory of Magnetoelectric Physics and Devices, State Key 	Laboratory of Optoelectronic Materials and Technologies, Institute of Neutron Science and Technology, School of Physics, Sun Yat-Sen University, Guangzhou, 510275, China
	}
	
	\author{Zhong-Yi Xie}
	\affiliation{Guangdong Provincial Key Laboratory of Magnetoelectric Physics and Devices, State Key 	Laboratory of Optoelectronic Materials and Technologies, Institute of Neutron Science and Technology, School of Physics, Sun Yat-Sen University, Guangzhou, 510275, China
	}
	
	\author{Cui-Qun Chen}
	\affiliation{Guangdong Provincial Key Laboratory of Magnetoelectric Physics and Devices, State Key 	Laboratory of Optoelectronic Materials and Technologies, Institute of Neutron Science and Technology, School of Physics, Sun Yat-Sen University, Guangzhou, 510275, China
	}

	\author{W\'ei W\'u}
	\email[Corresponding author: ]{wuwei69@mail.sysu.edu.cn}
	\affiliation{Guangdong Provincial Key Laboratory of Magnetoelectric Physics and Devices, State Key 	Laboratory of Optoelectronic Materials and Technologies, Institute of Neutron Science and Technology, School of Physics, Sun Yat-Sen University, Guangzhou, 510275, China
	}
	
	\author{Dao-Xin Yao}
	\email[Corresponding author: ]{yaodaox@mail.sysu.edu.cn}
	\affiliation{Guangdong Provincial Key Laboratory of Magnetoelectric Physics and Devices, State Key 	Laboratory of Optoelectronic Materials and Technologies, Institute of Neutron Science and Technology, School of Physics, Sun Yat-Sen University, Guangzhou, 510275, China
	}

	\date{\today}

	
	\begin{abstract}
		We perform a dynamical mean-field theory (DMFT) + Bethe-Salpeter equation (BSE) study of collective spin excitations in $\mathrm{La_3Ni_2O_7}$, using a bilayer two-orbital Kanamori-Hubbard model derived from density functional theory. Our DMFT calculations reveal a double spin-stripe ground state. On this basis, we apply the BSE to compute the lattice transverse spin susceptibility, which exhibits a damped spin-wave spectrum with a bandwidth of $\sim$0.1\ eV and a wave vector of ($\pi$/2, $\pi$/2), in good agreement with resonant inelastic x-ray scattering and inelastic neutron scattering data. We inspect the orbital-resolved bare susceptibility that provides insight into  the momentum dependence of the spin-wave line shape. Further analysis of the irreducible vertex  reveals notable frequency dependence in the $d_{z^2}$ orbital. 
		Finally, we investigate the evolution of the spectrum with varying Hund's coupling and electron fillings.
		Our study  bridges the  localized and itinerant descriptions of spin excitations in $\mathrm{La_3Ni_2O_7}$.
		
	\end{abstract}

	\keywords{Bilayer nickelates, Magnetic excitation, Dynamical mean-field theory, Bethe-Salpeter equation}
	
	\maketitle

	\section{\label{sec:1}Introduction}
	Magnetic correlations play a central role in high-temperature superconductivity. 
	In cuprates, superconductivity arises  from the staggered antiferromagnetic background which supports collective spin-wave excitations at half filling. 
	In the newly discovered bilayer nickelate superconductor $\mathrm{La_3Ni_2O_7}$~\cite{sun_signatures_2023}, resonant inelastic x-ray scattering (RIXS)~\cite{chen_electronic_2024,zhong_spin_2026,zhong_doping_2026} and inelastic neutron scattering (INS)~\cite{chen_single-stripe_2026} both reveal a common feature:  the low-energy spectrum below 0.1\ eV is characterized by a damped spin wave with a wave vector of $M/2$=($\pi$/2, $\pi$/2). A direct fit of the line shape to a Heisenberg model requires strong interlayer antiferromagnetic superexchange $J_\bot$~\cite{chen_electronic_2024,zhong_spin_2026,zhong_doping_2026,chen_single-stripe_2026,ni_spin_2025}, which is almost an order of magnitude larger than the in-plane counterparts, in stark contrast to the cuprates.  This large $J_\bot$ is increasingly believed to be the driving force of  superconductivity in the system~\cite{yyf2023prb,yyf2023prb2,sugang2024prl,wucj2024,ryee2024,sakakibara_possible_2024,zhangyh2023,luo_high-tc_2024,cpl_41_5_057403,gu_effective_2025,yf2026prb,maier_interlayer_2026,wang_spin_2026},  and is supported by growing experimental evidence~\cite{zhang_interlayer_2026,zhong_spin_2026,XIE20243221,jc35-gj2l}. 
	However, the local moment picture fails to capture the multi-orbital character in $\mathrm{La_3Ni_2O_7}$ and does not account for the substantial hole doping of the $d_{x^2-y^2}$ orbital, necessitating an improved multi-orbital description~\cite{luo_bilayer_2023}.

	Recently, a Hartree-Fock + random phase approximation (RPA) study based on a renormalized band model~\cite{mei_itinerant_2026} also reproduced the experimental spin-wave spectrum, and the authors argued that its magnetism is ``fundamentally itinerant'', favoring the opposite interpretation.
	Note that a direct connection is expected between the band structure and the collective spin excitation even in a strongly correlated compound, as the magnetic wave vector can often be traced to underlying Fermi-surface nesting~\cite{norman2000,mazin2008,johannes_microscopic_2009,mazin_pairing_2009,lane_competing_2023}.
	The RPA approach is also limited by its weak-coupling character and is valid only for small interactions~\cite{zhan_magnetic_2026}; otherwise, a renormalized band structure must be used~\cite{mei_itinerant_2026}. 
	
	The limitations of both the local-moment and itinerant pictures motivate a  dynamical mean-field theory (DMFT), which treats  kinetic energy and interaction on an equal footing. Previous DMFT studies (including cluster extensions) 
	have mainly focused on the orbital-selective renormalized electronic structure  as well as superconducting pairing~\cite{christiansson_correlated_2023,lechermann_electronic_2023,wu_superexchange_2024,yyf2024dmft,ouyang_hund_2024,yue_correlated_2025,26zoulj}. 
	However, a DMFT study of dynamical magnetic excitations is still missing and is essential for a unified picture of $\mathrm{La_3Ni_2O_7}$.

	Here, starting from a bilayer two-orbital Kanamori-Hubbard model, we perform a DMFT study of magnetic excitations in $\mathrm{La_3Ni_2O_7}$, combined with the Bethe-Salpeter equation (BSE). 
	Our computed spin susceptibility shows good agreement with the experimental spin wave spectrum, and we analyze the BSE to show how the momentum dependence of the spin wave is inherently associated with the bare susceptibilities and dynamical irreducible vertices.
	Our study provides a general perspective on the spin dynamics in $\mathrm{La_3Ni_2O_7}$.

	\section{\label{sec:mm}Model and Methods}
	The bilayer two-orbital tight-binding model is obtained from a Wannier fit to density functional theory results for bulk $\mathrm{La_3Ni_2O_7}$ at ambient pressure, where spin-density-wave order persists~\cite{liu_evidence_2022,meng_density-wave-like_2024,khasanov_pressure-enhanced_2025,zhao_pressure-enhanced_2025,plokhikh_unraveling_2025}.
	The model contains four orbitals (two Ni $e_g$ orbitals per layer), analogous to the pressurized case~\cite{luo_bilayer_2023} but with overall smaller hopping amplitudes. We also include a Kanamori interaction with Coulomb repulsion $U$ and Hund's coupling $J_H$. Details of the model can be found in Appendix~\ref{apx:model}.
	
	\begin{figure}
		\includegraphics[scale=0.8, trim=30 0 0 0]{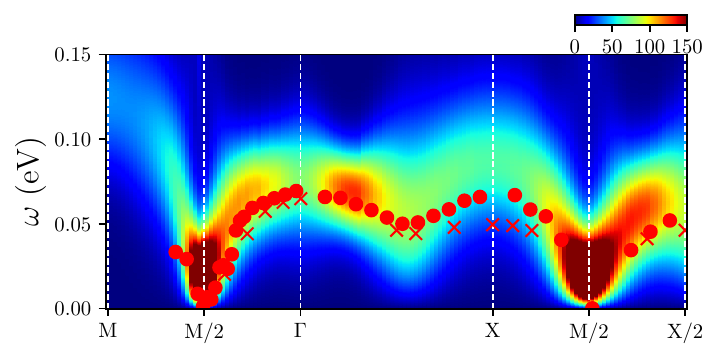}
		\caption{\label{fig:1} Imaginary part of the transverse spin susceptibility computed by DMFT+BSE in the odd channel of the DSS magnetic ground state: $\rm{Im}\chi_{\rm odd}^{\bot}(q,\omega)$. The fillings are $n_z$=1 and $n_x$=0.45.  $U=4.4$\ eV, $J_H/U=0.15$, $T$=1/60\ eV.  The red dots and crosses denote data from RIXS~\cite{chen_electronic_2024} and INS~\cite{chen_single-stripe_2026}, respectively.}
	\end{figure}
	
	The DMFT+BSE method follows a procedure similar to those in Refs.~\cite{mushoff_magnetic_2021,mushkaev_spin_2025,park_magnetic_2011,zhang_effect_2014}. 
	First, we perform the DMFT computation to determine the symmetry-broken magnetic ground state, using an eightfold supercell that matches the wave vector $M/2$. For the impurity problems, only local correlations within each Ni $e_g$ sector are retained, and each impurity is solved using the hybridization-expansion continuous-time quantum Monte Carlo method (CT-HYB)~\cite{wallerberger_w2dynamics_2019}.
	Our results reveal exclusively a double spin-stripe (DSS) configuration with no indication of single spin-stripe (SSS) order, which is unsurprising since mean-field solutions energetically favor spin polarization.
	Based on the DSS ground state, we again use CT-HYB to compute the  local two-particle Green's functions, and obtain the impurity irreducible vertices in the transverse spin particle-hole channel  via the local BSE.
	Next,  the impurity vertices are used to construct the lattice vertex $\Gamma(i\omega_n)$, in the same spirit as the local self-energy approximation.
	Finally, we substitute $\Gamma(i\omega_n)$ and the lattice bare susceptibility $\chi^0(q,i\omega_n)$ computed from fully interacting one-particle Green's functions, into the lattice BSE to obtain the dressed $\chi(q,i\omega_n)$. 
	Throughout this study, we use the full matrix form of the BSE without applying the density-density approximation (which retains only density-density-like terms relevant to the physical susceptibility and neglects the feedback from orbital-flip terms).
	We use a fermionic frequency box $N_{\nu}$=90, for which our tests indicate convergence.
	In the last step, we perform the orbital-frequency summation for $\chi(q,i\omega_n)$, and apply maximum-entropy analytic continuation~\cite{bergeron_algorithms_2016} to obtain the real-frequency spectra.
	Further methodological details are presented in Appendices~\ref{apx:bse} and \ref{apx:mm}.

	\section{\label{sec:result}Results}

	In Fig.~\ref{fig:1}, we present the computed imaginary part of the dynamical transverse spin susceptibility in the odd channel: $\rm{Im}\chi^\bot_{\rm odd}(q,\omega)$, with $U$=4.4\ eV, $J_H/U$=0.15, $T$=1/60\ eV. The fillings of the $d_{x^2-y^2}$ and $d_{z^2}$ orbitals are set to 0.45 and 1, respectively.  The result shows good agreement with the experimental spin wave spectrum indicated by red dots from RIXS~\cite{chen_electronic_2024} and crosses from INS~\cite{chen_single-stripe_2026}. The magnetic wave vector is clearly observed at $M/2$ as enforced by the Goldstone theorem. 
	Another important feature is the damping of the dispersion which closely resembles the RIXS result~\cite{chen_electronic_2024}. 
	For the susceptibility in the even channel, the intensity is much weaker than in the odd channel and can be neglected.

	\begin{figure}
		\includegraphics[scale=0.96, trim=10 0 0 0]{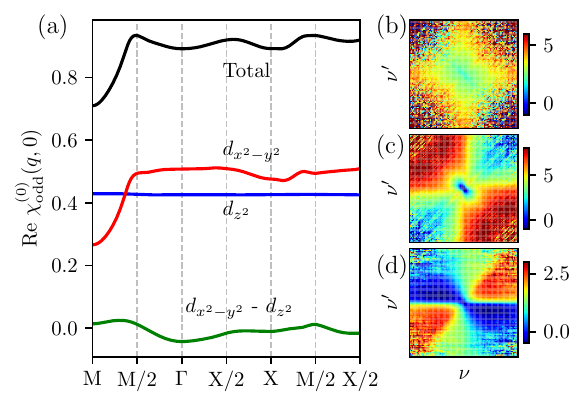}
		\caption{\label{fig:chi0qw}(a)  The real part of the orbital-resolved bare static susceptibilities in the odd channel: $\mathrm{Re} \chi^{0}_{\rm odd} (q,0)$. Red, blue, and green lines denote contributions from the $d_{x^2-y^2}$ orbital, the $d_{z^2}$ orbital, and interorbital terms, respectively. The black line denotes the total bare static susceptibility. (b-d) Irreducible-vertex components for the intra-$d_{x^2-y^2}$, intra-$d_{z^2}$, and interorbital channels, corresponding to onsite Coulomb interactions $U_x$, $U_z$, and $J_H$, respectively. The fermionic frequency box is $N_\nu$=90. }
	\end{figure}
	
	To understand the origin of this result, we analyze the lattice BSE and examine the real part of the orbital-resolved bare static susceptibilities $\rm{Re}\chi_{\alpha}^0(q,0)$. As shown  in Fig.~\ref{fig:chi0qw}(a),
	the black line denotes the total $\rm{Re}\chi^0(q,0)$, which can be decomposed into three contributions associated with the intraorbital $d_{x^2-y^2}$ (red), intraorbital $d_{z^2}$ (blue), and interorbital (green) terms. The momentum dependence of the total $\rm{Re}\chi^0(q,0)$ closely matches that of the spectrum in Fig.~\ref{fig:1}, where a larger magnitude of $\rm{Re}\chi^0(q,0)$ corresponds to a lower excitation energy in the spectrum. 
	The orbital-resolved $\rm{Re}\chi_{\alpha}^0(q,0)$ further shows that the peak at the wave vector $M/2$ arises mainly from intraorbital $d_{x^2-y^2}$ (red) and interorbital (green) components. 
	Specifically, the peak is produced by their sum, rather than by either contribution alone. 
	This feature can be better understood in terms of Fermi-surface nesting~\cite{botzel_theory_2024}:   the susceptibility in the odd channel is related to nesting of pockets with opposite mirror parities---here, the $\alpha$ and $\beta$ pockets. We further find that this nesting wave vector is mainly controlled by the $d_{x^2-y^2}$ orbital filling $n_x$. Within the DMFT approximation, $n_{x}$=0.45   places the peak near $M/2$ and also brings the Fermi surface into agreement with the angle-resolved photoemission spectroscopy result \cite{yang_orbital-dependent_2024}. For a larger $n_x$,  we will show in the following that the $\alpha$ pocket expands, causing the spectral peak to move toward the $M$ point.
	On the other hand, the orbital-resolved $\rm{Re}\chi_{\alpha}^0(q,0)$  reveals the orbital origin of the spectral shape along $\Gamma-X$ in which an enhancement appears near $\sim X/2$ relative to either side. Here, the left side ($X/2\rightarrow\Gamma$) is associated with a decrease in the interorbital component (green) while the right side ($X/2\rightarrow X$) is associated with the intra-$d_{x^2-y^2}$ component (red). 
	In contrast, the $d_{z^2}$ orbital shows a nearly featureless, flat contribution (blue), in line with strong localization at half-filling.

	The correspondence between the bare static susceptibility and the dynamical excitations in the above analysis may suggest an RPA enhancement scenario in which the vertex can be described by static screened interactions, i.e.,  the dynamical structure (frequency dependence) is unimportant.
	This is not the case. 
	In Fig.~\ref{fig:chi0qw}(b-d), we present the irreducible vertex in the particle-hole spin channel for representative components of (b) $d_{x^2-y^2}$, (c) $d_{z^2}$ orbitals, and (d) interorbital $d_{x^2-y^2}-d_{z^2}$ terms, corresponding to onsite Coulomb interactions $U_x$, $U_z$, and $J_H$, respectively. 
	The three vertices exhibit markedly different frequency dependences. 
	For the $d_{x^2-y^2}$ orbital in Fig.~\ref{fig:chi0qw}(b), aside from high-frequency noise, the vertex shows an almost uniform frequency structure with a weak but visible cancellation along $\nu$=$-\nu^\prime$.
	This is a typical feature of the single-band Hubbard model where the crossed particle-hole  spin and charge  channels largely cancel, leaving an almost constant background $\sim U_x$ together with a second-order particle-particle correction that depends on $\nu+\nu^\prime$~\cite{del_re_fluctuations_2021}. A simple average over the frequency box shows that $U_x\sim$3.8\ eV, which is smaller than the bare $U$=4.4\ eV and indicates appreciable screening in the $d_{x^2-y^2}$ orbital away from half-filling.
	However, for $U_z$ and $J_H$ in Fig.~\ref{fig:chi0qw}(c,d), the situation is quite different as they exhibit notable quadrant dependence, indicating a competition between crossed particle–hole and particle–particle channels~\cite{localelectronic21,del_re_fluctuations_2021}.
	This clearly demonstrates strong temporal fluctuations for the $d_{z^2}$ orbital, in stark contrast to its weak spatial fluctuations.
	Moreover, the central region in Fig.~\ref{fig:chi0qw}(c) is characterized by a near-zero structure elongated along $\nu$=-$\nu^\prime$, which closely resembles the structure in Ref.~\cite{kitatani_why_2019} associated with a substantial third-order particle-particle correction.
	
	\begin{figure}
		\includegraphics[scale=0.63, trim=20 0 0 0]{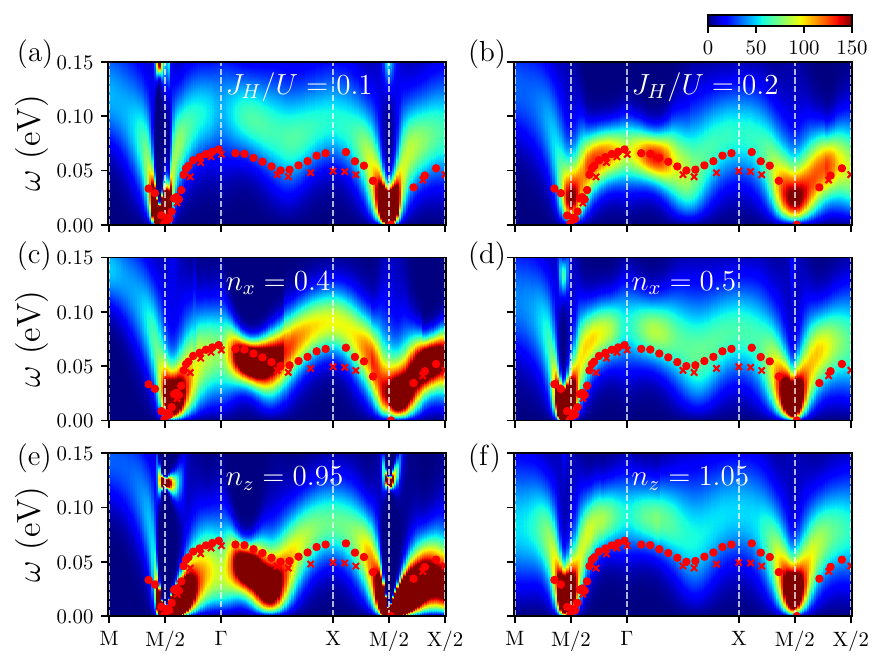}
		\caption{\label{fig:chiqw6} Imaginary part of the transverse spin susceptibility computed by DMFT+BSE  in the odd channel of the DSS magnetic ground state $\rm{Im}\chi_{\rm odd}^{\bot}(q,\omega)$ with varying (a,b) $J_H$, (c,d) $n_x$, and (e,f) $n_z$ with respect to Fig.~\ref{fig:1}.
			The red dots and crosses denote data from RIXS~\cite{chen_electronic_2024} and INS~\cite{chen_single-stripe_2026}, respectively.}
	\end{figure}

	We next examine how the spin-excitation spectrum evolves with Hund's coupling and electron fillings.
	Taking Fig.~\ref{fig:1} as a reference, in Fig.~\ref{fig:chiqw6} we show the results with varying (a,b) $J_H$, (c,d) $n_x$, and (e,f) $n_z$, respectively.
	Overall, these spectra generally maintain the original shape with changes occurring mainly near the $\Gamma$ point.
	For $J_H/U$=0.1 in Fig.~\ref{fig:chiqw6}(a), the dominant spectral weight shifts to approximately 0.1\ eV with weaker intensity, indicating substantially suppressed correlations. However, for $J_H/U$=0.2 in Fig.~\ref{fig:chiqw6}(b), the spectrum is nearly unchanged relative to $J_H/U$=0.15.
	For the changes in electron filling shown in Fig.~\ref{fig:chiqw6}(c-f), we find that the excitation near the $\Gamma$ point shifts to lower energy with notably enhanced intensity under hole doping, and vice versa under electron doping.
	Here, the spectrum is more sensitive to $n_z$ than to $n_x$, as demonstrated by the difference between the middle and bottom panels.
	Moreover, we find in Fig.~\ref{fig:chiqw6}(f) that electron doping of the $d_{z^2}$ orbital can notably broaden  the excitation at the $M$ point, an effect absent in the other cases.
	Experimentally, 
	a study of strain dependence in thin films also reports a visible change in the spin wave line shape near the $\Gamma$ point~\cite{zhong_spin_2026} consistent with our prediction.
	Another experiment revealed that  the spin wave is damped upon Sr doping~\cite{zhong_doping_2026}.
	However, given the uncertainty of the actual doping level and the purity  of samples, a quantitative comparison with our results remains open, and further experiments are needed to clarify this issue.

	\section{\label{sec:dis}Discussion}
	In this DMFT study, we do not consider the interlayer correlations which are generally included in previous cluster-DMFT studies~\cite{wu_superexchange_2024,ryee2024,yue_correlated_2025,xie_evolution_2025,mo_intertwined_2025}.
	In fact,  we have extensively tested the case in which all four orbitals are treated as a single impurity, but a paramagnetic (PM) ground state persists without any indication of spin-symmetry breaking. We consider that when the cluster size is small, the cluster-DMFT approximation may favor these short-range correlations. This is particularly evident in the dynamical cluster approximation study of the single-band Hubbard model~\cite{maier2005}, where antiferromagnetism (AFM) is suppressed by a local singlet and resonating valence bond state for cluster sizes $N_c$=2 and 4, respectively.
	On the other hand, the downfolding of the $d$-$p$ model of $\mathrm{La_3Ni_2O_7}$ to a four-band Ni-3$d$ orbital model may overestimate the inter-layer dimer state that competes with the magnetic order.
	A possible solution to this problem is to include in-plane nonlocal correlations that compete with interlayer correlations. However, that would make the impurity size  too large to be treated by a CT-HYB solver. Nevertheless, experience with the single-band Hubbard model indicates that including nonlocal correlations generally retains the physics predicted by single-site DMFT when the main focus is antiferromagnetism~\cite{mushoff_magnetic_2021}. Therefore, for a cluster DMFT solution in $\mathrm{La_3Ni_2O_7}$, we anticipate a similar result with a relatively lower N\'eel temperature,  consistent with the trend implied by the Mermin-Wagner theorem.

	This study further demonstrates that the magnetic excitations in both bilayer $\mathrm{La_3Ni_2O_7}$ and cuprates can be described within the same DMFT+BSE framework, despite their markedly different low-energy electronic structures, and  the spin bandwidth of the former is about one-third that of the latter. Specifically, Ref.~\cite{mushoff_magnetic_2021} shows that,  with $U$=4.7\ eV, $t$=0.4\ eV, and $|t^\prime/t|$=0.2 for a single-band Hubbard model, DMFT  can provide a good description of the spin waves in $\mathrm{La_2CuO_4}$ with a bandwidth of $\sim$0.25\ eV. In our model, we have nearly identical $t$=0.395\ eV, $|t^\prime/t|\approx$0.17 for the $d_{x^2-y^2}$ orbital,  and $U$=4.4\ eV, but the obtained spin wave bandwidth is $\lesssim$0.1\ eV. This indicates a significant reconstruction of the spin dynamics in $\mathrm{La_3Ni_2O_7}$ associated with two-orbital hybridization and Hund's coupling. 

	The multi-orbital effects lead us to another noteworthy question: why can two distinct pictures, the local moment~\cite{chen_electronic_2024,chen_single-stripe_2026} and itinerant~\cite{mei_itinerant_2026} descriptions, both provide a good description of the spin excitations in $\mathrm{La_3Ni_2O_7}$? For the local moment method, a Heisenberg model with two sites representing the two Ni cations per unit cell is commonly used, omitting orbital degrees of freedom and itinerancy, but the result remains plausible~\cite{chen_electronic_2024,chen_single-stripe_2026}; for the itinerant method, an RPA solution to a Hubbard model is also in reasonable agreement~\cite{mei_itinerant_2026}.
	To better address this question, in Appendices~\ref{apx:bsebnd} and \ref{apx:redbse}, we derive a reduced BSE that bridges the two existing pictures. The reduced BSE is obtained by reformulating the original BSE in the bonding representation in the PM phase and applying the density-density approximation, which retains the fermionic frequency dependence but has an orbital dimension of only 2$\times$2.
	Within this formulation, RPA is easily recovered by neglecting the frequency dependence of the vertices, yielding an explicit Stoner form of $\chi(q,\omega)$ showing how multi-orbital effects are involved (Appendix~\ref{apx:redbse}).
	However, we stress that this does not imply that the RPA vertices coincide with the zeroth-order background of the dynamical vertices,  since we have demonstrated that $\Gamma$ carries pronounced frequency structure extending to $\nu,\nu^\prime\rightarrow\infty$. 
	A phenomenological RPA description can only be recovered by summing over fermionic frequencies in the bare and dressed susceptibilities obtained from the BSE  and substituting both into the RPA equation: $\Gamma^{\rm RPA}=(\chi^0)^{-1}-(\chi^{\rm BSE})^{-1}$~\cite{Park2011}.
	For the local moment description, we point out that the superexchange scales are naturally encoded in the BSE~\cite{georges_dynamical_1996,szilva_quantitative_2023,stepanov_effective_2018,katanin_exchange_2023,katanin_exchange_2025} and can even be extracted    analytically in some cases~\cite{kiani_electronic_2016,mushoff_magnetic_2021}.
	In $\mathrm{La_3Ni_2O_7}$, DMFT+BSE extends the localized description by incorporating itinerancy in the strong-coupling regime, and reflected as the broadened spin-wave dispersion.  
	Our reduced BSE further justifies the  two-site setting in the Heisenberg model which is more naturally interpreted in the bonding representation, as is also emphasized in Ref.~\cite{mei_itinerant_2026}.

	\section{\label{sec:con}Conclusion}
	We have performed a DMFT+BSE study of magnetic excitations in  $\mathrm{La_3Ni_2O_7}$, using a bilayer two-orbital Kanamori-Hubbard model derived from a Wannier fit to density functional theory. 
	Our DMFT calculations reveal a robust double spin-stripe magnetic ground state.
	On this basis, we apply the BSE to investigate the magnetic excitation.
	With $U$=4.4\ eV, $J_H/U$=0.15, the computed lattice transverse spin susceptibility in the odd channel $\chi^\bot_{\rm odd}(q,\omega)$ reveals a damped spin wave spectrum with a bandwidth of about 0.1\ eV and a wave vector of $M/2$=($\pi$/2,$\pi$/2), in good agreement with the RIXS and INS data.
	Furthermore, we analyze the BSE to examine the orbital-resolved bare static susceptibilities $\chi^0_\alpha(q,0)$ and the irreducible vertices $\Gamma$,   
	where $\chi_\alpha^0(q,0)$ provide insight into  the momentum dependence of the spin-wave line shape,
	and $\Gamma$ exhibits notable frequency dependence that cannot be directly reduced to an RPA vertex.
	In addition, we investigate the evolution of the spin wave  with varying Hund's coupling $J_H$ and electron fillings $n_x$, $n_z$.
	Our study bridges the  two existing localized and itinerant scenarios of the spin excitation in $\mathrm{La_3Ni_2O_7}$, and  further supports the DMFT+BSE framework in the study of collective spin dynamics in multiorbital systems.

	\begin{acknowledgments}
		This work was supported by National Natural Science Foundation of China (12494591, 12494594, 92565303), National Key Research and Development Program of China (2022YFA1402802), Guangdong Provincial Quantum Science Strategic Initiative (GDZX2601002 and GDZX2401010), Guangdong Provincial Key Laboratory of Magnetoelectric Physics and Devices (2026B1212170049), Research Center for Magnetoelectric Physics of Guangdong Province (2024B0303390001), and Center for Advanced Study of SYSU. We also acknowledge the Dali Nickelate Forum.
	\end{acknowledgments}

	\appendix
	
	\section{\label{apx:model}Bilayer two-orbital Kanamori-Hubbard model}
	
	The bilayer two-orbital Kanamori-Hubbard model is expressed as $H=H_0+H_U$, with
	\begin{align}
		H_0&=\sum_{\rm k\sigma} \Phi^\dag_{\rm k\sigma} H({\rm k}) \Phi_{\rm k\sigma},\\
		H({\rm k})&=\left(\begin{array}{cc}
			H_{||}({\rm k}) & H_{\bot}({\rm k})\\
			H_{\bot}({\rm k}) & H_{||}({\rm k})
		\end{array}\right),\nonumber \\
		H_{||}({\rm k})&=\left(\begin{array}{cc}
			T_{{\rm k}}^{z} & V_{{\rm k}}\\
			V_{{\rm k}} & T_{{\rm k}}^{x}
		\end{array}\right),\ H_{\bot}({\rm k})=\left(\begin{array}{cc}
			t_{\bot}^{z} & V_{{\rm k}}^{\prime}\\
			V_{{\rm k}}^{\prime} & t_{\bot}^{x}
		\end{array}\right), \nonumber\\
		T_{{\rm k}}^{z/x}&	=2t_{1}^{z/x}(\cos k_{x}+\cos k_{y})+4t_{2}^{z/x}\cos k_{x}\cos k_{y}+\epsilon^{z/x}, \nonumber \\
		V_{{\rm k}}&	=2t_{3}^{xz}(\cos k_{x}-\cos k_{y}),\ V_{{\rm k}}^{\prime}=2t_{4}^{xz}(\cos k_{x}-\cos k_{y}), \nonumber
	\end{align}
	and
	\begin{align}\label{eq:kanamori}
		H_U=&U\sum_{i\alpha} n_{i\alpha\uparrow} n_{i\alpha\downarrow} + U^\prime \sum_{i\sigma,\alpha<\beta} n_{i\alpha\sigma} n_{i\beta\bar{\sigma}}\\
		+&(U^\prime-J_H) \sum_{i\sigma,\alpha<\beta}   n_{i\alpha\sigma} n_{i\beta\sigma}\nonumber\\
		+&J_H^\prime \sum_{i,\alpha\neq\beta} ( c^\dag_{i\alpha\uparrow} c^\dag_{i\alpha\downarrow} c_{i\beta\downarrow} c_{i\beta\uparrow}- c^\dag_{i\alpha\uparrow} c_{i\alpha\downarrow} c^\dag_{i\beta\downarrow} c_{i\beta\uparrow} ). \nonumber
	\end{align}

	Here, $H_0$ is the tight-binding Hamiltonian in the basis $\Phi=(c_{1z},c_{1x},c_{2z},c_{2x})^T$. $c_{iz}/c_{ix}$ denotes the annihilation operator for the Ni $d_{z^2}/d_{x^2-y^2}$ orbital on the $i$th layer.
	The hopping parameters are shown in Table~\ref{tab:hop1}. For comparison, we also list the hopping parameters from Ref.~\cite{luo_bilayer_2023} for the high-pressure case in parentheses. 
	
	For $H_U$, a Kanamori interaction is used. Besides the Coulomb repulsion $U$ and Hund's coupling $J_H$, $U^\prime$=$U-2J_H$, $J_H^\prime=J_H$ are taken. Note that in Eq.~\ref{eq:kanamori} the sum over orbital indices $\alpha,\beta$ runs only within each Ni-$e_g$ sector. Other notation follows standard conventions.

	\begin{table}[h]
		\begin{tabular}{ccccc}
			\hline \hline 
			
			$t_{1}^{x}$ & $t_{1}^{z}$ & $t_{2}^{x}$ & $t_{2}^{z}$ & $t_{3}^{xz}$  \tabularnewline\hline 
			-0.395 & -0.076 & 0.065 & -0.013 & 0.188 
			\tabularnewline
			(-0.483) & (-0.110) & (0.069) & (-0.017) & (0.239)
			\tabularnewline \hline \hline
			
			$t_{\bot}^{x}$ & $t_{\bot}^{z}$ & $t_{4}^{xz}$ & $\epsilon^{x}$ & $\epsilon^{z}$  \tabularnewline \hline 
			0.014 & -0.537 & -0.027 & 0.839 & 0.432 
			\tabularnewline
			(0.005) & (-0.635) & (-0.034) & (0.776) & (0.409)
			\tabularnewline  
			
			\hline \hline
		\end{tabular}
		\caption{\label{tab:hop1}Tight-binding  parameters  for bulk $\mathrm{La_3Ni_2O_7}$ at ambient pressure.  $\epsilon^{x},\epsilon^z$ are onsite energies for Ni-$d_{x^2-y^2}, d_{3z^2-r^2}$ orbitals, respectively.    The values in parentheses are taken from~\cite{luo_bilayer_2023} and correspond to the high-pressure case. All values are in eV.
		}
	\end{table}

	\section{\label{apx:bse}BSE convention}
	We use the convention for the two-particle Green's function defined in Ref.~\cite{localelectronic21},
	\begin{align}
		G^2_{\alpha\beta\gamma\delta}&=\langle\mathcal{T} c^\dag_\alpha c_\beta  c^\dag_\gamma c_\delta \rangle,
	\end{align}
	where $\alpha,\beta,\gamma,\delta$ denote the composite spin-orbital  indices. Throughout this study we only consider the transverse channel with fixed  spin indices $\uparrow\downarrow\downarrow\uparrow$. 
	
	The bare particle-hole susceptibilities for the impurity and lattice are expressed respectively as
	\begin{align}
		\label{eq:chi0}
		\chi^{0\nu\nu^{\prime}\omega}_{\alpha\beta\gamma\delta}&=-\beta\delta_{\nu\nu^{\prime}} G^{imp}_{\delta\alpha}(\nu)G^{imp}_{\beta\gamma}(\nu+\omega),\\
		\chi^{0\nu\nu^{\prime}\omega}_{\alpha\beta\gamma\delta}(q)&=-\frac{\beta\delta_{\nu\nu^{\prime}} }{N_k} \sum_{k} G_{\delta\alpha}(k,\nu)G_{\beta\gamma}(k+q, \nu+\omega).\nonumber
	\end{align}

	The BSE is expressed in the spin (magnetic) channel as
	\begin{align}
		\bar{\chi}_{m}&=\left(I-\frac{1}{\beta^{2}}\bar{\chi}^{0}\bar{\Gamma}_{m}\right)^{-1}\bar{\chi}^{0}. \label{eq:bse}
	\end{align}
	Here we adopt a minus sign in the BSE kernel so that $\Gamma_m$ is proportional to $U$. We also use a bar to denote the matrix form of a quantity, which is related to its tensor form by
	\begin{align}\label{eq:t2m}
		O^{\nu\nu^\prime\omega}_{\alpha\beta\gamma\delta} = \bar{O}^\omega|_{(\nu\alpha\beta),(\nu^\prime\delta\gamma)}.
	\end{align}
	
	The BSE matrix therefore has dimension  $N_\nu N_{orb}^2$, with $N_\nu$,  $N_{orb}$ denoting the size of the fermionic frequency box and the number of orbitals, respectively. In this study, a maximum of $N_\nu$=90 is used, for which our tests indicate convergence.

	\section{BSE versus RPA}
	To compare the BSE with RPA, we explicitly derive their relation. We show that by dropping the $\nu,\nu^\prime$ dependence of the vertex, the BSE reduces to the RPA form. The proof is straightforward using the ladder form of Eq.~\ref{eq:bse}
	\begin{align}
		\bar{\chi}&=\bar{\chi}^{0}+\frac{1}{\beta^{2}}\bar{\chi}^{0}\bar{\Gamma}\bar{\chi} \nonumber\\
		&=\bar{\chi}^{0}\left[1+\frac{1}{\beta^{2}}\bar{\Gamma}\bar{\chi}^{0}+\frac{1}{\beta^{4}}\bar{\Gamma}\bar{\chi}^{0}\bar{\Gamma}\bar{\chi}^{0}+...\right]. \label{eq:bseladder}
	\end{align}
	
	When $\Gamma$ is independent of $\nu,\nu^\prime$,  we have  $\bar{\Gamma}_{\nu\nu^\prime}=U$, thus
	\begin{align}
		[\bar{\Gamma}\bar{\chi}^0]_{\nu\nu^\prime}&=U\chi^{0\nu^\prime}, \nonumber \\ [\bar{\Gamma}\bar{\chi}^0\bar{\Gamma}\bar{\chi}^0]_{\nu\nu^\prime}&=\beta^2U^2\chi^{0\omega}\chi^{0\nu^\prime}, \nonumber\\
		&...\nonumber
	\end{align}
	with $\chi^{0\omega}=\frac{1}{\beta^2}\sum_{\nu\nu^\prime} \bar{\chi}^0_{\nu\nu^\prime}=\frac{1}{\beta^2}\sum_\nu\chi^{0\nu}$.
	Substituting these expressions into Eq.~\ref{eq:bseladder}, we get
	\begin{align}
		\bar{\chi}_{\nu\nu^{\prime}}&=\chi^{0\nu}\left[\delta_{\nu\nu^{\prime}}+\frac{1}{\beta^{2}}U\chi^{0\nu^{\prime}}+\frac{1}{\beta^{2}}U^{2}\chi^{0\omega}\chi^{0\nu^{\prime}}+...\right] \nonumber
	\end{align}
	
	Applying frequency summation $\frac{1}{\beta^2}\sum_{\nu\nu^\prime}$ on both sides of the above equality, we recover the RPA equation
	\begin{align}
		\chi^\omega&=\chi^{0\omega}\left[1+U\chi^{0\omega}+(U\chi^{0\omega})^{2}+...\right] \nonumber\\
		&=\frac{\chi^{0\omega}}{1-U\chi^{0\omega}}.
	\end{align}
	
	
	\section{\label{apx:bsebnd}BSE in the bonding representation in the PM phase}
	In this section, we derive the BSE in the bonding representation for $\mathrm{La_3Ni_2O_7}$. Since the fermionic frequency and orbital indices act on different degrees of freedom, we suppress $\nu,\nu^\prime$ in the following. 
	
	We first express
	\begin{align}
		\chi^{0}=\sum_{\lambda} \hat{\phi}_{\lambda}^{\dag}  \bar{\chi}^{0}_\lambda \hat{\phi_{\lambda}}  \label{eq:chi0bnd},
	\end{align} 
	with the basis defined as
	\begin{align}
		\hat{\phi}_{\lambda}^{\dag}&=(c_{z\lambda_1}^{\dag}c_{z\lambda_2},c_{z\lambda_1}^{\dag}c_{x\lambda_2},c_{x\lambda_1}^{\dag}c_{z\lambda_2},c_{x\lambda_1}^{\dag}c_{x\lambda_2}), \\
		\hat{\phi}_{\lambda}&=(c_{z\lambda_2}^{\dag}c_{z\lambda_1},c_{x\lambda_2}^{\dag}c_{z\lambda_1},c_{z\lambda_2}^{\dag}c_{x\lambda_1},c_{x\lambda_2}^{\dag}c_{x\lambda_1})^T. 
	\end{align}
	Here $\lambda=(\lambda_1,\lambda_2)$, $\lambda_{1/2}=\pm$, yielding a total of 4 blocks in Eq.~(\ref{eq:chi0bnd}). However, this does not imply that the BSE also separates into four blocks. According to $Z_2$ symmetry, we group (++,--) and (+-,-+) into the even (+) and odd (-) channels, respectively, i.e.,
	\begin{align}
		\bar{\chi}_{+}^{0}={\rm{diag}(\bar{\chi}_{++}^{0},\bar{\chi}_{--}^{0})},\  \bar{\chi}_{-}^{0}={\rm{diag}(\bar{\chi}_{+-}^{0},\bar{\chi}_{-+}^{0})}. \label{eq:chi0pm}
	\end{align}
	
	The irreducible vertex is similarly obtained by a bonding/antibonding rotation at the two-particle level:
	\begin{align}\label{eq:Gammapm}
		\bar{\Gamma}_\pm&=\frac{1}{2}\left(\begin{array}{cc}
			1 & 1\\
			1 & 1
		\end{array}\right)\otimes \\
		&\left(\begin{array}{cccc}
			U_{z} &  &  & J_{H} \\
			& U^{\prime}-J_{H}+J_{H}^{\prime} & J_{H}^{\prime}\\
			& J_{H}^{\prime} & U^{\prime}-J_{H}+J_{H}^{\prime}\\
			J_{H} &  &  & U_{x}
		\end{array}\right).\nonumber
	\end{align}
	These matrix elements retain their dependence on $\nu,\nu^\prime$, while the symbols denote their RPA counterparts.
	Also note that we distinguish the intraorbital components for  $d_{z^2}$ and $d_{x^2-y^2}$ orbitals by $U_z$ and  $U_x$, respectively, since they are significantly different in DMFT.
	
	Similarly, the BSE in the bonding representation becomes
	\begin{align}
		\bar{\chi}_{\pm}&	=[I-\frac{1}{\beta^2}\bar{\chi}_{\pm}^{0}\bar{\Gamma}_{\pm}]^{-1}\bar{\chi}_{\pm}^{0}, \label{eq:bsebnd}
	\end{align}
	and the orbital sum  is
	\begin{align}
		\chi_{\rm even/odd} &= \sum_{\alpha\beta\in(1,4,5,8)} \bar{\chi}_{\pm\alpha\beta}. \label{eq:chieo}
	\end{align}

	\section{\label{apx:redbse}Reduced BSE in the PM phase}
	In this section, we  show that the 8$\times$8 BSE in Eq.~\ref{eq:bsebnd} can be  reduced to only 2$\times$2, by applying the density-density approximation. 
	First, according to the definition of the physical susceptibilities in Eq.~\ref{eq:chieo}, we drop all the irrelevant non-density-density terms from Eq.~\ref{eq:bsebnd}, yielding a 4$\times$4 form
	
	\begin{align}
		\bar{\chi}^\prime_{\pm}&	=\bar{L}^\prime_{\pm}\bar{\chi}_{\pm}^{0\prime} \label{eq:chip}, \\
		\bar{L}^\prime_{\pm}&	=[I-\frac{1}{\beta^2}\bar{\chi}_{\pm}^{0\prime}\bar{\Gamma}^\prime_{\pm}]^{-1}, \label{eq:Lp}
	\end{align}
	with
	\begin{align}
		\bar{\Gamma}_{\pm}^{\prime}&=\frac{1}{2}\left(\begin{array}{cc}
			1 & 1\\
			1 & 1
		\end{array}\right)\otimes\bar{\Gamma}^{red},\\ 
		\bar{\Gamma}^{red}&=\left(\begin{array}{cc} \label{eq:Gammared}
			U_{z} & J_{H}\\
			J_{H} & U_{x}
		\end{array}\right),\\
		\bar{\chi}_{+}^{0\prime}&={{\rm diag}(\bar{\chi}_{++}^{0\prime},\bar{\chi}_{--}^{0\prime})},\label{eq:chi0eveneff} \\ \bar{\chi}_{-}^{0\prime}&={{\rm diag}(\bar{\chi}_{+-}^{0\prime},\bar{\chi}_{-+}^{0\prime})}, \label{eq:chi0oddeff}
	\end{align}
	\begin{align}
		\bar{\chi}_{\lambda_1\lambda_2}^{0\prime}&	=\left(\begin{array}{cc}
			\chi_{z\lambda_1z\lambda_2z\lambda_2z\lambda_1}^{0} & \chi_{z\lambda_1z\lambda_2x\lambda_2x\lambda_1}^{0}\\
			\chi_{x\lambda_1x\lambda_2z\lambda_2z\lambda_1}^{0} & \chi_{x\lambda_1x\lambda_2x\lambda_2x\lambda_1}^{0}
		\end{array}\right). \label{eq:chi0eop}
	\end{align}
	
	In this case, Eq.~\ref{eq:chieo} simply becomes
	\begin{align}
		\chi_{\rm even/odd}=\sum_{\alpha\beta} \bar{\chi}^\prime_{\pm\alpha\beta}. \label{eq:chieo2}
	\end{align}
	
	Next, observing that Eqs.~\ref{eq:chi0eveneff}-\ref{eq:chi0oddeff} remain in block-diagonal form, we can rewrite  Eq.~\ref{eq:Lp} as
	\begin{align}
		\bar{L}_\pm^\prime = I+ \frac{1}{2\beta^2}\bar{\chi}^{0\prime}_\pm 
		\left( 
		\begin{array}{cc}
			\bar{\Gamma}^{red}\bar{L}^{red}_\pm & \bar{\Gamma}^{red}\bar{L}^{red}_\pm \\
			\bar{\Gamma}^{red}\bar{L}^{red}_\pm & \bar{\Gamma}^{red}\bar{L}^{red}_\pm
		\end{array} \right), \label{eq:Lp2}
	\end{align} 
	with the definitions
	\begin{align}
		\bar{L}_\pm^{red} &= (I-\frac{1}{\beta^2}\bar{X}^0_\pm \bar{\Gamma}^{red})^{-1},\label{eq:Lred} \\
		\bar{X}^0_\pm&=\frac{1}{2}(\bar{\chi}_{+\pm}^{0\prime}+\bar{\chi}_{-\mp}^{0\prime}).
	\end{align}
	
	By substituting Eq.~(\ref{eq:Lp2}) into Eq.~(\ref{eq:chip}) and applying $\frac{1}{2}\sum_{\alpha\beta}$ on both sides, we obtain a 2$\times$2 matrix form of the reduced BSE
	\begin{align}
		\bar{X}_\pm = \bar{L}^{red}_\pm \bar{X}^0_\pm \label{eq:bseeff}.
	\end{align}
	
	Now Eq.~(\ref{eq:chieo2}) becomes
	\begin{align}
		\chi_{\rm even/odd} =2 \sum_{\alpha\beta}\bar{X}_{\pm\alpha\beta}. \label{eq:chieo3}
	\end{align}
	
	The only approximation in the above derivation is the density-density approximation; all remaining steps are exact matrix operations.
	We have tested the density-density approximation, and find it highly accurate in the PM phase. Moreover, for momenta along the diagonal direction, it becomes exact  since the hybridization vanishes.

	Lastly, we drop the fermionic frequency dependence of Eq.~\ref{eq:Gammared} to obtain an explicit form of  Eq.~\ref{eq:chieo3}.
	Expanding Eq.~\ref{eq:Lred}   as
	\begin{align}
		\bar{L}_\pm^{red}&=\frac{1}{\Delta}\left(\begin{array}{cc}
			1-d & b\\
			c & 1-a
		\end{array}\right), \label{eq:Lpexp}\\
		\Delta&=(1-a)(1-d)-bc,\nonumber
	\end{align}
	with the shorthand notation
	\begin{align}
		a:=zU_{z}+yJ_{H}\nonumber,&\quad
		b:=zJ_{H}+yU_{x},\label{eq:shorthand}\\
		c:=yU_{z}+xJ_{H},&\quad
		d:=yJ_{H}+xU_{x}, \\
		\frac{1}{\beta^2}\sum_{\nu\nu^\prime} \bar{X}^0_\pm:&=\left(\begin{array}{cc}
			z & y\nonumber\\
			y & x
		\end{array}\right).
		\nonumber
	\end{align}
	
	With these definitions, Eq.~\ref{eq:chieo3} is derived as
	\begin{widetext}
		\begin{align}
			\chi_{\rm even/odd}&=\frac{2}{\Delta}\left[
			x+z+2y-(xz-y^2)(U_x+U_z-2J_H)
			\right] \label{eq:chieo4},\\
			\Delta&=1-(zU_{z}+xU_{x}+2yJ_{H})+(xz-y^{2})(U_{z}U_{x}-J_{H}^{2}). \label{eq:Delta}
		\end{align}
	\end{widetext}
	
	The denominator exhibits a Stoner form with the first bracket corresponding to the leading Stoner enhancement  while the second bracket denotes a second-order correction that tends to weaken the former.

		\section{\label{apx:mm}DMFT+BSE Computations in the AFM phase}
		
		\begin{figure}
			\includegraphics[scale=0.4]{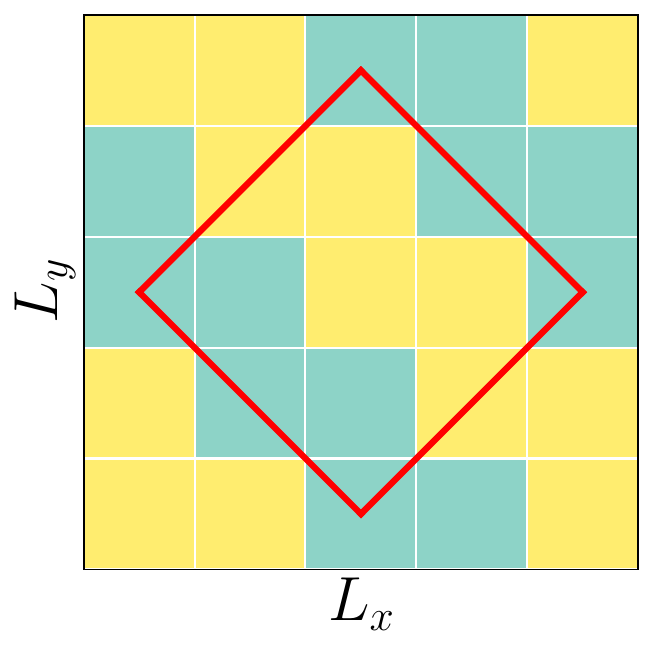}
			\caption{\label{fig:patch} Schematic of the magnetic supercell in DMFT+BSE. Each patch denotes a unit cell of the bilayer two-orbital model. The red diamond indicates an eightfold supercell with a total of 32 orbitals, which accommodates both the SSS and DSS configurations.  }
			
		\end{figure}
		
		To account for magnetic order with wave vector $M/2$=($\pi$/2,$\pi$/2),  we adopt an eightfold supercell, as illustrated  in Fig.~\ref{fig:patch} by the red diamond. This supercell accommodates both the SSS and DSS configurations.  Here, each patch denotes a unit-cell containing four orbitals from the bilayer two-orbital model, leading to a total $N_{orb}$=32 in the supercell.
		The DMFT impurities are constructed within each Ni-$e_g$ shell, i.e., only local correlations are retained, which results in 16 independent impurity problems during each DMFT iteration.
		
		With these settings, our calculations reveal an exclusive DSS magnetic ground state regardless of the choice of the finite $U$, $J_H$.
		Further inspection shows that these 16 impurity problems reduce to two inequivalent impurity problems, associated with the upper and the lower layers in the light blue patch (or the lower and upper layers in the yellow patch)  in Fig.~\ref{fig:patch}.
		This is equivalent to adopting a fourfold supercell~\cite{hjp24prb}, which substantially improves the efficiency of the DMFT and BSE calculations.
		
		For the impurity solver, we use the hybridization-expansion continuous-time quantum Monte Carlo method (CT-HYB) implemented in $\textsc{w2dynamics}$~\cite{wallerberger_w2dynamics_2019}. The  single-particle Green's function is computed in the Legendre basis, which is cross-checked using the $\textsc{triqs/cthyb}$~\cite{seth_triqscthyb_2016} package. The two-particle Green's function is computed using worm sampling, and is also cross-checked using exact diagonalization implemented in $\textsc{pomerol}$~\cite{Krivenko2025POMEROL}.
		
		In the BSE, the matrix size is $1024N_{\nu}$, according to Eq.~\ref{eq:t2m} and Fig.~\ref{fig:patch}. For a practical $N_\nu\sim100$, the matrix size would reach $\sim 10^5$, which makes direct inversion computationally demanding.
		In fact, we can take full advantage of the lattice symmetry and the locality of the vertex to reduce the matrix size to only 16$N_{\nu}$, which  is achieved as follows.
		
		First, as shown in Fig.~\ref{fig:patch}, the double stripe pattern contains reflection about the $x=y$ axis and a permutation between the upper/lower layer in one sublattice (light blue) and the lower/upper layer in the other (yellow). These two symmetries separate the full BSE into four smaller blocks of size 256$N_\nu$.
		In this case, we rotate both $\bar{\chi}^0$ and $\bar{\Gamma}$, and rotate the resulting $\bar{\chi}^{rot,i}$ ($i$=1$\sim$4) back to the orbital space after solving all four BSE blocks. 
		Note that, for  $\bar{\chi}^{0}$, we construct it directly from the rotated single-particle Green's function, using Eq.~\ref{eq:chi0}.
		For $\bar{\Gamma}$, given its local structure, the rotation can also be performed efficiently in sparse form.
		
		Next, for each block $\bar{\Gamma}^{rot,i}$, we observe that it remains highly  symmetric with many repeated rows/columns, because it is constructed from the local vertices of two inequivalent impurities. By exploiting this redundancy, we can reduce the vertex to a smaller subspace and recover its full representation through
		\begin{align}\label{eq:bsevsp}
			\bar{\Gamma}^{rot,i}=\bar{P} \bar{\gamma} \bar{P}^\dagger.
		\end{align}
		Here $\bar{\gamma}$ is a $16N_{\nu}\times16N_{\nu}$ matrix, and $\bar{P}$ is $256N_{\nu}\times16N_{\nu}$. Using the Woodbury matrix identity, we can therefore rewrite the block BSE as
		\begin{align}
			\bar{\chi}&=\bar{\chi}^0 (I+\frac{1}{\beta^2} \bar{P} \bar{\gamma}\bar{\kappa}),\\
			\bar{\kappa}&= (I-\frac{1}{\beta^2} \bar{P}^\dagger \bar{\chi}^0 \bar{P} \bar{\gamma})^{-1} \bar{P}^\dagger \bar{\chi}^0.
		\end{align}
		The block index is suppressed for simplicity.

		\bibliography{ref2}

	\end{document}